\documentstyle[preprint,aps]{revtex}
\begin{document}
\draft
\begin{titlepage}
\title{Resonance flourescence in atomic coherent systems:
Spectral features.}
\author{Sandhya S.N.}    
\address{The Mehta Research Institute for Mathematics and Mathematical Physics\\
Chatnag Road Jhusi, Allahabad 211019 India\\}

\date{}

\maketitle

\begin{abstract}
We study resonance flourescence in a four level ladder system and illustrate
some novel features due to quantum interference and atomic coherence
effects. 
We find that under three photon resonant conditions, in some region
of the parameter space of the rabi frequencies $\Omega_1,\Omega_2,\Omega_3$ ,
emission is dominantly by the level 4 at the line center 
 even though there is an almost equal distribution  
of populations in all the levels.  As one increases $\Omega_3$
with $\Omega_1 {\rm and}\Omega_2$ held fixed, the four level
system 'dynamically collapses' to a two level system. The steady state
populations and the the resonance flourescence from all the levels
provide adequate evidence to this effect. 

{\bf PACS(numbers)}: 42.50.Hz., 32.80
\end{abstract}

\end{titlepage}

Four-level atomic systems with
three photon resonant interaction, lead to very different phenomena 
than those due to two photon effects in three level systems \cite{arimondo},
as revealed by  the absorption characteristics in these systems \cite{sand}.
Such characteristics have been attributed to quantum interference (QI)
and atomic coherent (ACh) effects.
The interplay amongst the driving field strengths and
detunings  control the nature of QI and ACh mechanisms
which in turn form a powerful tool for controlling the spontaneous emission and
flourescence from various levels of the atomic system \cite{lit,knight}.
To recapitulate a few known
results, while the occurence of dark resonances\cite{arimondo} and electromagnetically
induced transparency (EIT)\cite{jea} were typical
effects expected in two photon processes  , in 4-level systems
interacting with three driving fields, narrow absorption at the
line center was reported \cite{sand}. Other four level studies
have revealed as varied a phenomena like photon switching \cite{harris}, 
two photon absorption and inhibition \cite {harsha} and other 
interfence effects\cite{lukin}. 
Effect of QI on the suppression of spontaneous emission \cite{hui},
coherent control of polarization of light \cite{wileandy}
have been demonstrated experimentally and
the effect of  relative
phases of the lasers \cite{peng} has also been reported.

In this communication we study the
resonance flouresence from various levels of a 4-level atomic system
in the parameter space
spanned by the rabi frequencies of the driving fields , $(\Omega_1,
\Omega_2,\Omega_3)$. Apart from the already known features like 
supression of emission , narrowing of emission line and
Mollow splitting of lines at high driving intensities,
we present novel features exhibited 
 in certain regions in this parameter
space. 
For  low \cite{foot}  ground state excitation energies, we find that
the four level system 'dynamically collapses' to a two level system for certain values
of $ \Omega_2$  and $\Omega_3$. That is, two of the levels get 'dynamically
decoupled' from the rest of the system.
Further we identify the region in the parameter space where we can 
control the flourescence from the 4th level.
A look at the steady state populations in these regions
further throws light on the nature of the interferences.
This rich variety of phenomena 
may find
possible applications in 
the areas of
high resolution, control of flouresence at definite frequencies,
quantum computing \cite{prl}  and population transfer\cite{hioe}.

As already mentioned, the model we consider is a 
 4-level ladder system driven by three resonant fields.
 Here (see fig 1), 1-2, 2-3 and 3-4 are the allowed
dipole transitions. 
 We take the decay
contants of the system to be very similar to that of the Rubidium
system of reference \cite{jea}. The energy level separations are denoted by
$ \omega_{21},\omega_{32} $ and $\omega_{43}$ respectively.
We sketch the outlines of the calculation here, the details 
will be provided elsewhere.
In setting up the hamiltonian for the four level system, 
we treat the driving fields classically
and write it in the standard form

\begin{eqnarray}
{\cal H}  &=& \sum_{i=1}^4 {\cal E}_i a_i^{\dagger} a_i +
         \sum_{j=1}^3 {\hbar} \Omega_j ( e^{-i \omega_j t} a_{j+1}^{\dagger}a_j + {\rm h.c.})
\label{eq1}
\end{eqnarray}
where $\Omega_j=\frac {\mu_{j,j+1}.E_j}{\hbar}$ is the rabi frequency of the
coupling fields , $a^{\dagger}_i,a_i$
describe the creation and annihilation of the electrons in 
the respective levels i,
and ${\cal E}_i$ are the energy levels of the atomic system in the
absence of any external field. 
We follow the approach of Agarwal\cite{agarwal} and Narducci et. al.\cite{scully}, 
for obtaining the flourescence spectrum.
We employ the equation of motion of the density matrix
of the atomic system in the interaction representation,

\begin{equation}
\frac{\partial \rho}{\partial t}= - \frac{i}{h} [{\cal H}_I,\rho]+ {\cal L}_{irrev}
\label{eq2}
\end{equation}
where ${\cal H}_I$ is the hamiltonian in the interaction respresentaion,
and $ {\cal L}_{irrev}$ is the irreversible part of the Liouville operator
described by the master equation given in reference \cite{agarprofopt}. 


The density matrix equations (\ref{eq2})  are rewritten in the rotating wave approximation
as


\begin{eqnarray}
\frac {\partial {\bar \rho}_{12}}{\partial t} &=& (i \Delta_1 -\Gamma_{2}/2)
{\bar \rho_{12}} -i \Omega_1 ({\bar \rho_{22}}-{\bar \rho_{11}}) + i \Omega_2
{\bar \rho_{13}} \nonumber \\ 
\frac {\partial {\bar \rho}_{23}}{\partial t} &=& (i \Delta_2 -(\Gamma_{2}+\Gamma_3)/2)
{\bar \rho_{23}} -i \Omega_1 {\bar \rho_{13}}-i \Omega_2 ({\bar \rho_{33}}-{\bar \rho_{22}}) +
 i \Omega_3 {\bar \rho_{24}} \nonumber \\
\frac {\partial {\bar \rho}_{34}}{\partial t} &=& (i \Delta_3 -(\Gamma_{3}+\Gamma_4)/2)
{\bar \rho_{34}} -i \Omega_2 {\bar \rho_{24}}-i \Omega_3 ({\bar \rho_{44}}-{\bar \rho_{33}}) 
 \nonumber \\
\frac {\partial {\bar \rho}_{13}}{\partial t} &=& (i( \Delta_1+\Delta_2) -\Gamma_{3}/2)
{\bar \rho_{13}} -i \Omega_1 {\bar \rho_{23}}+i \Omega_2 {\bar \rho_{12}}+
i \Omega_3 {\bar \rho_{14}}) \nonumber \\
\frac {\partial {\bar \rho}_{14}}{\partial t} &=& (i( \Delta_1+\Delta_2+\Delta_3) -\Gamma_{4}/2)
{\bar \rho_{14}} -i \Omega_1 {\bar \rho_{24}}+i \Omega_3 {\bar \rho_{13}}
 \nonumber \\
\frac {\partial {\bar \rho}_{24}}{\partial t} &=& (i( \Delta_2+\Delta_3) -(\Gamma_{2}+\Gamma_4)/2)
{\bar \rho_{24}} -i \Omega_1 {\bar \rho_{14}}-i \Omega_2 {\bar \rho_{34}}+
i \Omega_3 {\bar \rho_{23}}) \nonumber \\
\frac {\partial {\bar \rho}_{22}}{\partial t} &=&  -\Gamma_{2} {\bar \rho_{22}}+
i \Omega_1 ({\bar \rho_{21}}-{\bar \rho_{12}}) + i \Omega_2
({\bar \rho_{23}}-{\bar \rho_{32}})+ \gamma_{23} {\bar \rho_{33}} + \gamma_{24} 
{\bar \rho_{44}} \nonumber \\
\frac {\partial {\bar \rho}_{33}}{\partial t} &=&  -\Gamma_{3} {\bar \rho_{33}}+
i \Omega_3 ({\bar \rho_{34}}-{\bar \rho_{43}}) - i \Omega_2
({\bar \rho_{23}}-{\bar \rho_{32}})+ \gamma_{34} {\bar \rho_{44}}  
 \nonumber \\
\frac {\partial {\bar \rho}_{44}}{\partial t} &=&  -\Gamma_{4} {\bar \rho_{44}}-
i \Omega_4 ({\bar \rho_{34}}-{\bar \rho_{43}})  
\label{spc}
\end{eqnarray}
where $\gamma_{ik}$ are the transition rates from the k to ith levels,
 $ \Gamma_i$ are the decay constants of the levels i and
$\Delta_i=\omega_{i,i+1}-\omega_i$ are the laser detunings
(${\bar \rho_{ij}} = {\bar \rho_{ji}}^{*}$ and ${\rm Tr}{\bar \rho}=1).$
The level shifts 
and pure phase relaxations have been ignored  in our analysis of the spectrum. 
We also ignore the contribution from $\gamma_{ik}$ for $i > k$ which corresponds
to an excitation to a higher level. In other words we assume that
the energy level separations  to be far greater than the thermal excitation energy.
Since it is easier to evaluate the solutions in the Laplace space ,
we recast (\ref{spc}) in a more compact form as

\begin{equation}
\frac {d \psi}{d t}= M \psi +C
\label{eq5}
\end{equation}
where we identify $ \psi_1={\bar \rho_{12}},\psi_2={\bar \rho_{23}},\psi_3={\bar \rho_{34}},
\psi_4={\bar \rho_{13}},\psi_5={\bar \rho_{14}},\psi_6={\bar \rho_{24}},
\psi_7={\bar \rho_{22}},\psi_8={\bar \rho_{33}},\psi_9={\bar \rho_{44}},
{\rm and} \psi_{i+9}=\psi_i^*$ , M is a $(15 \times 15)$ matrix , 
and $C$  is the inhomogenous part with the elements $C_1=C_{10}^*=-i \Omega_1$
and the rest being zero. By taking the Laplace transform of the above 
equation one obtains the solution in the simple form

\begin{equation}
{\bar \psi_{i}}(z)=\sum_j {\cal M}_{ij}(z) \psi_j(\tau_0) +
\frac {1}{z} \sum_j {\cal M}_{ij}(z) C_j.
\label{eq6}
\end{equation}
where ${\bar \psi_i}(z)$ is the Laplace transform of 
$\psi (t)$ and  ${\cal M}= (z-M)^{-1}$. The steady state solution is obtained
by taking the limit $z \rightarrow 0 ({t \rightarrow \infty}) $in the above equation.


The flouresence spectrum is proportional to the steady state 
correlation function  of the scattered field,  $ \lim_{t \rightarrow \infty} 
< E^{-}({\bf r},t+\tau) E^{+} ({\bf r},t)> $,  where $E^-,E^+$ correspond to the
positive and negative frequency part of the scattered radiation. 
Since the source field
operators and the atomic polarization operators are directly proportional
to each other \cite{spring} , we can express the flouresence spectrum in terms of the
atomic correlation functions as 

\begin{equation}
\Gamma^1(z) =\int_{0}^{\infty} d \tau_1 \lim_{t \rightarrow \infty}
<{\cal P}^{\dagger} (t+\tau_1) {\cal P}(t)> e^{-i \omega t}.
\label{eq7}
\end{equation}
where the atomic polarization operator ${\cal P}$ is given by

\begin{equation}
{\cal P}^{\dagger}= \sum_{i=1}^{3}\mu_{i,i+1} a_{i+1}^{\dagger} a_{i}.
\label{eq8}
\end{equation}   
where $\mu_{ij}$ are modulus of the induced dipole moments.

Applying the quantum regression theorem 
the steady
state spectrum is obtained, after some algebra,  as 

\begin{equation}
<{\cal P}^{\dagger} (z) {\cal P} (\infty)>=
\sum_{i=1}^3 (\mu_{i,i+1}^2 ({\cal M}_{ii} (z_i) \psi_{i+6}(\infty) +
 \sum_j (\frac {1}{z_i}
{\cal M}_{ij} C_j )\psi_i^*(\infty))
\label{eq9}
\end{equation}
where we have dropped all the rapidly oscillating terms.
Here $z_i=z-i\omega_i , z = i \omega$. Thus the spectrum has three distinct parts corresponding
to the center frequencies at $ \omega_i,i=1,3 $. Subtracting the delta 
function contributions which corresponds to the coherent part of the spectrum, 
we get the real part of the incoherent contribution to be

\begin{equation}
\Gamma^{(1)}_{incoh} =
\sum_{i=1}^3 (\mu_{i,i+1}^2 ({\cal M}_{ii} (z_i) \psi_{i+6}(\infty) +
\sum_j( \frac {1}{z_i}
{\cal N}_{ij} (z_i)C_j) \psi_i^*(\infty))
\label{eq10}
\end{equation}
where ${\cal N}=M^{-1} {\cal M}$.  With this expression for the 
correlation funcion, we proceed  to
numerically analyse the spectral features in the various domains
of the parameter space. 

We shall assume, for the sake of convenience, 
that the induced dipole moments of all the
three transitions are equal and look at the emission spectrum at three
photon resonance $\Delta_i=0,i=1,3$. Further we write all the parameters
$\Delta_i,\Gamma_i,\gamma_{ik}$ and $\Omega_i$ in terms of $\gamma$  
and set $\gamma=1$ .
The emission spectrum of the 4 level system is presented in two regions
of the parameter space centered around the points, $(7 ,4 ,1)$, (fig2(a)) and 
$( 7, 4, 50 )$ (fig2(b).  They show some new and
interesting features. We find in the former region that,  
at the line center,  emission is dominated  by the $4 \rightarrow 3$ transition
 and the other 
two emissions are relatively small. 
A look at the variation of the peak intensities as one transits from low to high
$\Omega_3$ indicates (fig4) that
the emission of the $4 \rightarrow 3$ transition 
peaks around $2\gamma$ and thereafter falls rapidly.
Emission for the  $2 \rightarrow 1$ and $3 \rightarrow 2$ is
 rather weak in this region.
A look at the population distributions (fig3(a)), at this peak value,
however reveals that at zero detuning , $\rho_{22}(\infty) \approx \rho_{44}(\infty)$, 
and $ \rho_{33}(\infty)$ is larger by 10 \%. 
This implies that in this region, the spectral features are controlled more by 
the atomic coherent effects 
than the population dynamics.

We now move over to the other region around (7, 4, 50) in 
the parameter space and the features
are given in fig3. The flourescence in this region is dominated by level 2 
and there is a total suppression of emission from levels 3 and 
4 (see fig2(b) ). 
 The populations in the levels
(fig 3(b)) also seem to indicate that the system, now,
effectively behaves  like 
a two level
system driven by a single resonant field. 
This is further substantiated by results in fig4
which shows the saturation of the peak intensity of emission
from level 2 with increase in $\Omega_3$, while  the peak intensities
corresponding to levels 3 and 4 show a rapid fall and approaches
zero.
This can be understood
hueristically by arguing that as $ \Omega_3$ is increased, the 
two photon 
induced coherence between the levels 2 and 4 creates an 'EIT' like
situation, which inhibits absorption from level 2 and 
the  population is forced
to reside in the level 2. 

To conclude, we have seen that various competing atomic
coherent and quantum interference effects are responsible for the control
of flourescence from various levels which might provide
a powerful tool
for such switching mechanisms. We have also seen the possibility
of a four level system behaving effectively like a two level system.
This feature might possibly be exploited in suppressing multi-photon
processes when there is a  need to work with high ground state
excitation energies. A study of the
photon statistics and collective effects in such systems, which will be  dealt
with else where, should reveal any difference in the
usual two level system and the 'dynamical' two level system.

\pagebreak

{\bf Figure Captions}

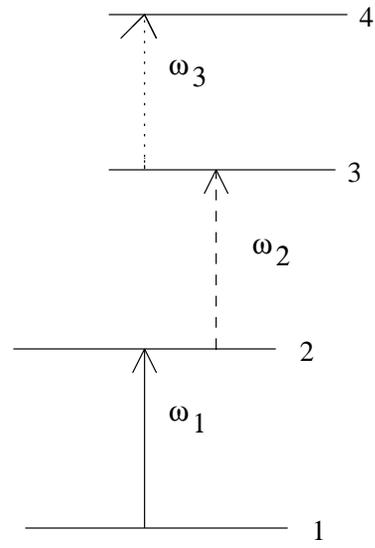
{\bf Figure1:}

Schematic energy level diagram of the four level system. $\omega_1,
\omega_2 {\rm and} \omega_3$ are the frequencies of the driving fields.
\\ \\
{\bf Figure 2:}

Power spectrum of the radiation emitted 
in the regions of driving field strenghts ${\bf a)}$ (7,4,1) and
${\bf b)}$ (7,4,50). 
The parameters are 
$\Gamma_2=6.0,\Gamma_3=\Gamma_4=1.0,\gamma_{23}=\gamma_{34}=1.0,
{\rm and} \gamma_{24}=0.0$, $\Delta_1= \Delta_2 = \Delta_3 =0.$
For easy comparison 
we have superposed the  centers of all the emission lines.
\\ \\
{\bf Figure 3:}

Steady state populations in the levels 2,3 and 4 corresponding to the 
regions in the parameter space ${\bf a)}$ (7,4,1) and
${\bf b)}$ (7,4,50) as a function of $\Delta_1$. The rest of the 
parameters are the same as in fig2.
\\ \\
{\bf Figure 4:}

Peak intensity of the emitted field corresponding to the transitions
 $4 \rightarrow 3$ , $3 \rightarrow 2$ and  $2 \rightarrow 1$
as a function of the driving field strength $\Omega_3$. The rest
of the parameters same as in fig2(a).


\begin{references}
\bibitem{foot}The resulting features are not very different for higher
excitation energies of the ground state. In fact the only difference
is in the appearance of the ususal Mollow sidebands. This will be reported
else where.
\end{references}
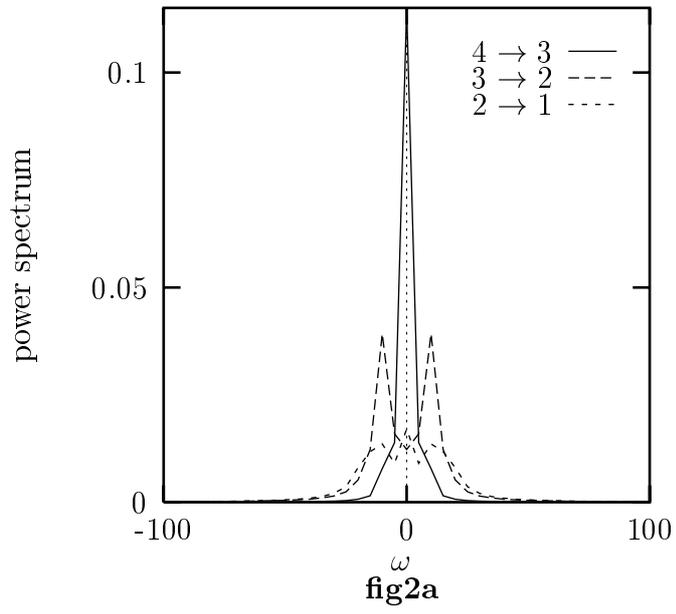
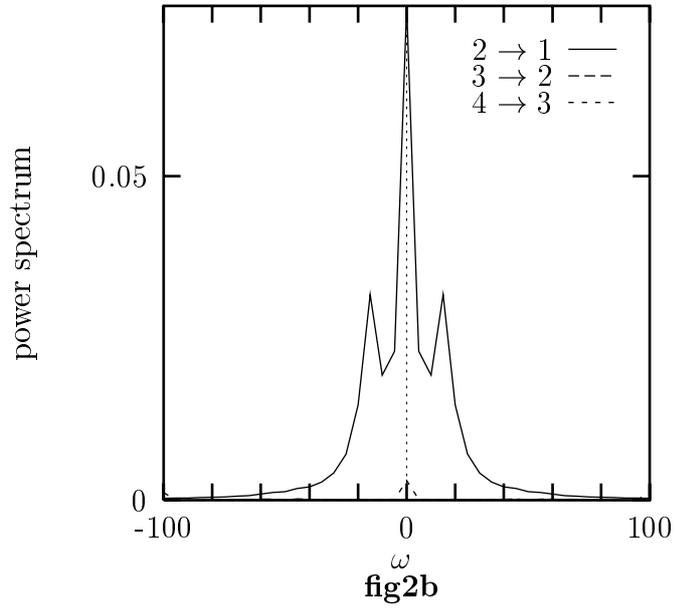
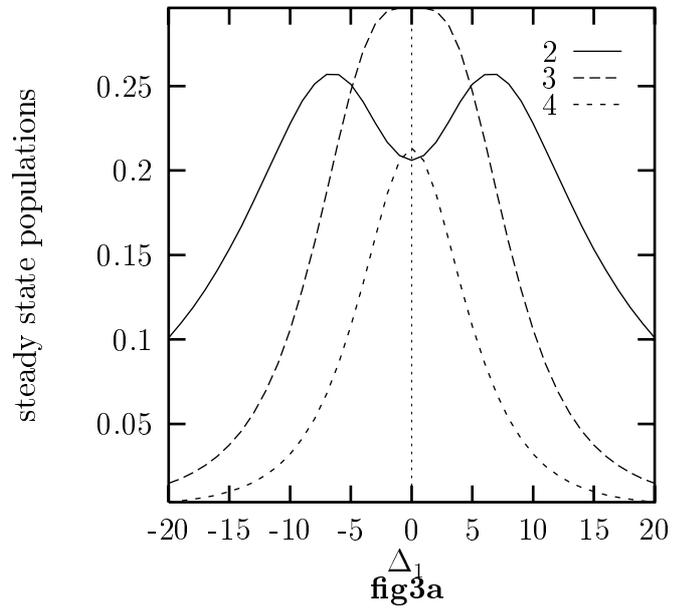
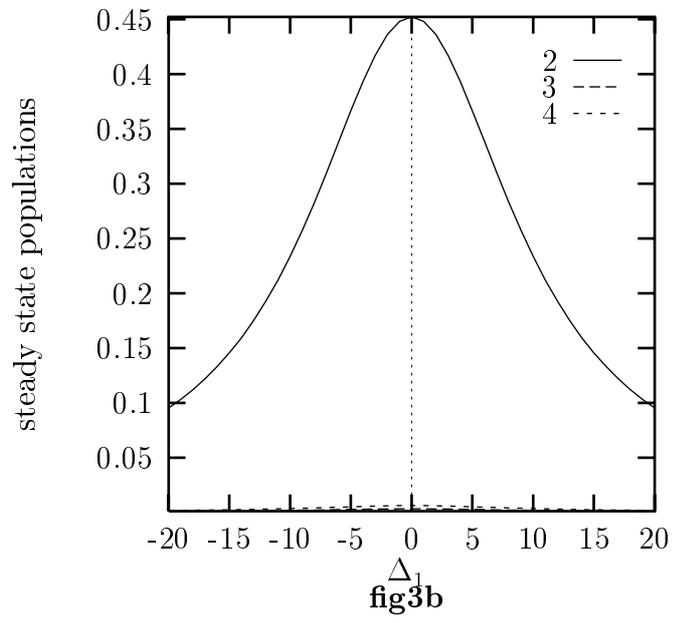
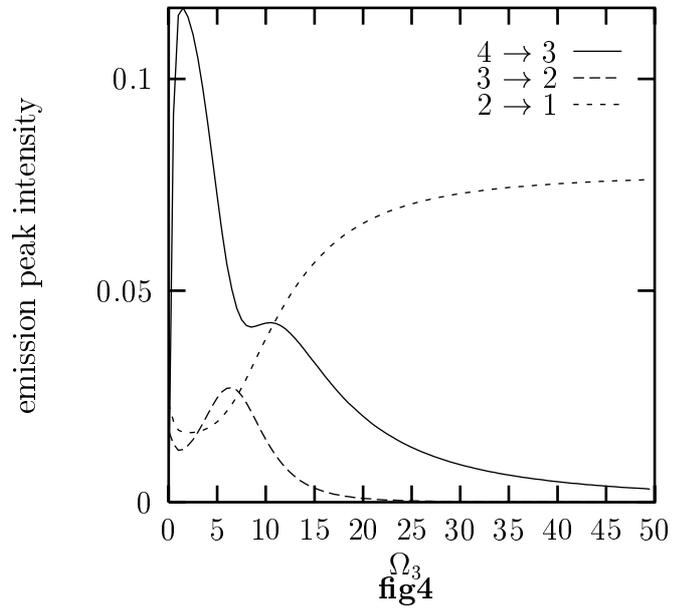
\end{document}